\documentclass[lettersize,journal]{IEEEtran}
\usepackage{tcolorbox}
\usepackage[numbers,sort&compress]{natbib}
\usepackage{balance}
\usepackage{amssymb,amsmath}
\usepackage{graphicx}
\usepackage{algorithm}
\usepackage[noend]{algorithmic} 
\usepackage{colortbl,booktabs}
\usepackage{multirow}
\usepackage{float}
\usepackage{xspace}
\usepackage{subfig}
\usepackage{setspace}
\newcommand{\scheme}{\textsc{WAFBooster}\xspace}

\newcounter{codefigure}
\usepackage{tikz}
\newcommand*{\circled}[1]{\lower.7ex\hbox{\tikz\draw (0pt, 0pt)%
		circle (.5em) node {\makebox[1em][c]{\small #1}};}}

\usepackage{hyperref}
\usepackage[hyphenbreaks]{breakurl}

\newenvironment{inditemize}{
	\begin{list}{$\bullet$}{
			\setlength{\labelwidth}{-6pt}
			\setlength{\itemsep}{0pt}
			\setlength{\leftmargin}{\labelwidth}
			\addtolength{\leftmargin}{\labelsep}
			\setlength{\parindent}{0pt}
			\setlength{\listparindent}{\parindent}
			\setlength{\parsep}{0pt}
			\setlength{\topsep}{0pt}}}{\end{list}}

\begin{document}

\title{\scheme: Automatic Boosting of WAF Security Against Mutated Malicious Payloads}

\author{
	\small Cong Wu$^1$, Jing Chen$^1$$^{*}$, Simeng Zhu$^1$, Wenqi Feng$^1$, Ruiying Du$^1$, and Yang Xiang$^2$\\
	\emph{$^1$ Wuhan University, China.}
	\emph{$^2$ Swinburne University of Technology, Australia.}
	\texttt{\{cnacwu,chenjing,zhusim,fengwq,duraying\}@whu.edu.sg} \qquad \texttt{yxiang@swin.edu.au}
	\thanks{$^{*}$ Corresponding author: Jing Chen.}}
\maketitle

\begin{abstract}
	Web application firewall (WAF) examines malicious traffic to and from a web application via a set of security rules. It plays a significant role in securing Web applications against web attacks. However, as web attacks grow in sophistication, it is becoming increasingly difficult for WAFs to block the mutated malicious payloads designed to bypass their defenses. In response to this critical security issue, we have developed a novel learning-based framework called \scheme, designed to unveil potential bypasses in WAF detections and suggest rules to fortify their security.
	Using a combination of shadow models and payload generation techniques, we can identify malicious payloads and remove or modify them as needed. \scheme generates signatures for these malicious payloads using advanced clustering and regular expression matching techniques to repair any security gaps we uncover. In our comprehensive evaluation of eight real-world WAFs, \scheme improved the true rejection rate of mutated malicious payloads from 21\% to 96\%, with no false rejections. \scheme achieves  a false acceptance rate 3$\times$ lower than state-of-the-art methods for generating malicious payloads. With  \scheme, we have taken a step forward in securing web applications against the ever-evolving threats.
\end{abstract}


\section{Introduction}

Web applications have been extensively deployed on the internet, enabling numerous convenient services such as webmail and online banking~\cite{li2011survey}. These applications, however, face significant threats in the form of web attacks, including command injection and cross-site scripting. Recent statistics reveal that web application attacks account for 32\% of all threats, making them the most pervasive form of attack activity~\cite{NTTreport}. To mitigate these risks, WAFs act as a protective shield, analyzing HTTP traffic to and from web applications and deciding whether to forward it to the application or block it~\cite{waf}.

Traditional WAFs rely on predefined security policies that are complex and require expert security knowledge for administration, such as Modsecurity~\cite{ModSecurity}. However, as web application defense and attack strategies evolve, web attack payloads are continually altered to bypass the security rules of WAFs~\cite{li2022evolutionary}. To stay up-to-date with the latest web attack payloads, the security community shares information on the top web application security flaws and malicious traffic flows that match security rules, through organizations such as the Open Web Application Security Project (OWASP)~\cite{OWASPTen}.

Despite relying on rule-based pattern matching detection techniques and extensive rule databases, traditional firewalls are struggling to keep up with the ever-evolving and mutating attack payloads, which are specifically designed to bypass their defenses. As a result, the objective is to strengthen the security of existing WAFs by proactively detecting and preventing the evasion of unknown malicious payloads. By doing so, the WAF can provide more efficient and effective protection against evolving threats, ensuring the safety and reliability of web applications for users.

The critical step in detecting unknown attack payloads against WAFs is to generate sufficient testing inputs to evade the security rules of WAFs. Existing studies fall into two categories, namely mutation-based methods~\cite{lee2020fuse,demetrio2020waf,Appelt2014} and generation-based methods~\cite{appelt2018machine,Wang2017,Lyu2019}. Mutation-based methods focus on altering existing payloads by applying carefully designed mutation schemes to derive testing inputs. In contrast, generation-based methods aim to design payload generation strategies and grammars to generate testing inputs from scratch. Nevertheless, these methods depend heavily on human expert knowledge of specific attacks to design mutation schemes or generation strategies. Additionally, the designed mutation schemes or generation strategies are dependent on specific attacks and cannot be generalized to other attacks. For instance, mutation schemes or generation strategies for command injection attacks may not work for cross-site scripting attacks. As a result, existing methods are low in efficiency and ineffective in enhancing WAF security.

\textbf{Our approach.}
In this paper, we present a novel black-box learning-based automatic WAF security boosting framework, called \scheme, which is the first of its kind. Our fundamental insight is that payloads mutated from detected malicious inputs against a specific WAF have a high likelihood of retaining malicious characteristics, even if the WAF may not immediately recognize them.
Thus, \scheme automatically exposes the underlying unknown malicious payloads against the given WAF and learns their signatures to enhance the security against of unknown malicious payloads.

To achieve this, \scheme creates a shadow model that imitates the target WAF's behavior, using a given black-box WAF.
WAFBooster operates in several key steps to enhance WAF security. Initially, it identifies malicious payloads that have been detected previously. From these, it generates new payloads specifically designed to bypass the shadow model. The next step involves payload correction, where the system alters any generated malicious payloads into valid formats and eliminates any payloads deemed invalid. Finally, WAFBooster categorizes the malicious payloads. Based on this categorization, it generates signatures for these payloads. These signatures are then used to update the WAF's security policies, thereby bolstering its defense mechanisms.
While many existing methods are limited to specific attacks and necessitate in-depth expert knowledge, \scheme offers a more flexible approach. Though it utilizes individual shadow models tailored to each attack type, the foundational RNN architecture remains consistent. This design allows \scheme to be adaptable, accommodating various attack payloads by adjusting model parameters without the need for extensive architectural modifications.

Nonetheless, there are two significant challenges to address. Firstly, it is difficult to expose both valid and malicious payloads. This means that the generated payloads should not only bypass the WAF but also conform to the grammar of attack scripts to be executed. Secondly, it is also challenging to produce an effective signature for the generated malicious payloads without increasing false alarms, i.e., the benign payloads blocked by the WAF.

To address the first challenge, we employ shadow training techniques to train a shadow model that imitates the behavior of the given real WAF. We use RNN for the payload generation to produce mutated malicious payloads. To ensure that the payloads are valid, we incorporate a payload corrector that measures the semantic similarity between the generated and original payloads. The corrector removes invalid payloads or modifies them to be valid.
Regarding the second challenge, we first design four score functions to identify important substrings in the payloads within a large search space. We then cluster the substrings using the metric of edit distance. We also create the efficient matching score-based regular expression on malicious payloads to produce the simplest signatures. These techniques enable us to produce effective signatures for the generated malicious payloads while minimizing false alarms.

In summary, we make the following contributions:
\begin{inditemize}
	\item We propose \scheme, the first effective automatic WAF security boosting framework. It constructs the shadow model, generates malicious payloads against the shadow model, ensures that the generated payloads are valid, and provides signatures of malicious payloads. It does not require any expert knowledge of specific attacks and can be extended to support other attack payloads.

	\item  To tackle the problem of generating valid and malicious payloads, we utilize the technique of \emph{shadow training} to construct the shadow model (\S~\ref{sec:shadow}). Furthermore, we have devised a neural network-based approach for generating malicious payloads that is based on \emph{sequence generation} (\S~\ref{sec:payload}). Additionally, we have developed a \emph{payload corrector} that measures the semantic similarity between the generated and original payloads, which ensures the validity of the payload (\S~\ref{sec:corr}).

	\item  To address the problem of generating precise signatures to update the security rules of WAF,
	we cluster the substrings within payload strings using the edit distance on important substrings.
	We design an efficient regular expression-based approach, designed to maximize performance within the context of our application(\S~\ref{sec:signature}).

	\item Our evaluation across eight WAFs revealed that \scheme increased the true rejection rate for mutated malicious payloads from 21\% to 96\%, with a false rejection rate of 0 (\S~\ref{sub:signature}). Furthermore, \scheme achieved a false acceptance rate three times lower than leading methods for generating malicious payloads (\S~\ref{sub:comparison}), underscoring its efficacy and efficiency in bolstering WAF security.

\end{inditemize}

\section{Background}
\label{sec:Background}

This section briefs web application attacks and adversarial examples.
\subsection{Web Application Attacks}
Web applications are an essential and widely used means of delivering information and services over the internet. As most web applications interact with backend database systems that store users' sensitive information, they have become the targets of various web attacks and are inherently vulnerable to exploitation. Examples of these attacks include SQL injection, cross-site scripting, cookie theft, session riding, browser hijacking, and others~\cite{NTTreport}. Breaches in web applications can lead to significant economic losses, ethical and legal consequences, and an enormous amount of information being compromised~\cite{webatt,yang2024pkt,sun2023privacy}.
To protect web applications from such attacks, WAFs are used. This work focuses on security against three primary types of attacks: SQL injection, cross-site scripting, and command injection.

\begin{inditemize}
	\item \textbf{SQL injection:} This attack allows an attacker to manipulate the queries that an application sends to its database~\cite{Appelt2014}. By inserting malicious SQL statements into an entry field, the attacker can interfere with the application's queries and manipulate the SQL query structure to execute unintended queries.
	\item \textbf{Cross-site scripting:} In this type of attack, the attacker manipulates a vulnerable web application to return malicious scripts to users. Since the web browser interprets all web responses returned by the trusted web application, the attacker can fully compromise the interaction with the application if the malicious scripts execute inside a victim's browser.
	\item \textbf{Command injection:} This attack allows the attacker to execute arbitrary commands on the web server and fully compromise the application and all its data. It is possible when an application passes unsafe user-supplied data, such as forms, cookies, or HTTP headers, to a system shell. Exploiting trust relationships, the attacker can pivot the attack to other systems within the organization by leveraging the command injection vulnerability to compromise the hosting infrastructure.
\end{inditemize}
\subsection{Adversarial Examples}

Deep learning has significantly impacted various applications~\cite{fang2024ic3m,yuan2024satsense,lin2022channel,zhang2025lcfed,lin2022tracking,zhang2024satfed,lin2024fedsn} but are susceptible to minor, human-imperceptible perturbations, leading to misclassification of adversarial examples~\cite{nowroozi2022demystifying,wucong2024tifs,yang2023efficient,carlini2017towards,li2024privacy}. This poses risks in security-sensitive areas like malware detection~\cite{grosse2017adversarial} and network intrusion detection~\cite{yang2018adversarial}, and biometric authentications~\cite{wu2021toward,tang2024merit,wu2020caiauth,wu2019icauth}, highlighting the need for a deeper security-focused understanding of neural networks.
Our approach enhances WAF security by employing RNNs to generate new malicious payloads, pushing the WAF's detection capabilities to their limits, similar to adversarial testing principles. Starting with initial tokens like `SELECT' for SQL injections, the RNN systematically generates attack-specific sequences. Considering the WAF as a binary classifier for network traffic, the application of adversarial techniques, well-explored in computer vision~\cite{carlini2017towards,nowroozi2022demystifying}, to the generation of structured adversarial payloads presents a novel challenge in improving WAF security.

\section{Design of \scheme}
\label{sec:design}
This section presents overview and detailed design.

\subsection{Overview}
A payload refers to a specific input designed to exploit vulnerabilities in web applications or data transmitted between a client and a server. These can be classified into two categories: benign and malicious payloads. \emph{Benign payloads} are routine data that users input into applications without intent to exploit or harm the system, such as entering a legitimate username like \texttt{username123} in a login field. \emph{Malicious payloads}, on the other hand, are crafted to exploit potential vulnerabilities, such as inputting \texttt{username123'; DROP TABLE users} to exploit SQL vulnerabilities. Malicious payloads can be single strings with multiple words, commands, or complex patterns like SQL commands or JavaScript snippets for SQL injections and XSS attacks. Throughout this paper, tasks like the scoring function refer to evaluating or processing these individual payloads based on their unique characteristics.

\begin{figure}[!t]
	\centering
	\includegraphics[width=.8\linewidth]{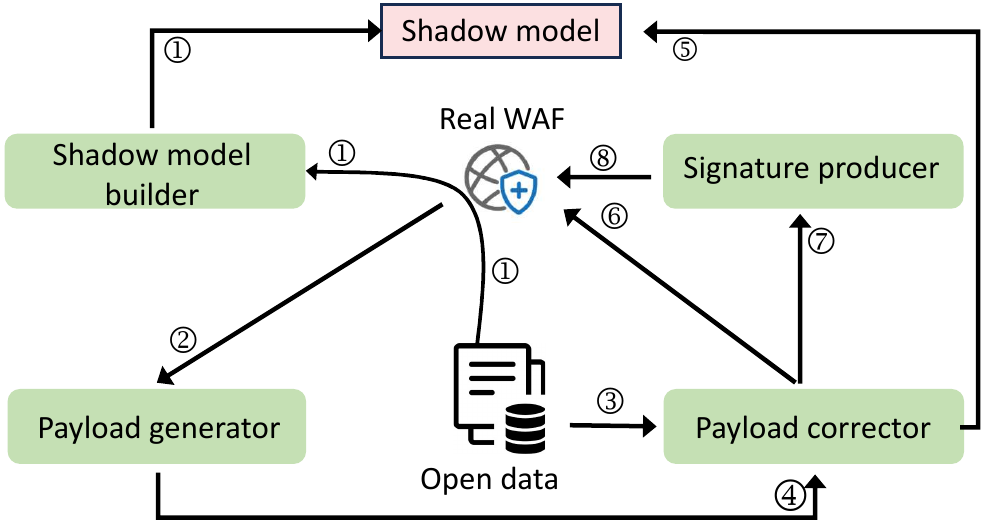}
	\caption{Workflow of \scheme}
	\label{Fig.overview}
\end{figure}

To enhance the identification and detection of previously unknown malicious payloads, a novel approach called \scheme has been developed. This approach generates corresponding signatures capable of matching with malicious payloads, which enables the vulnerabilities of a given WAF to be addressed. As depicted in Fig.~\ref{Fig.overview}, \scheme comprises four main modules: the \emph{shadow model builder}, the \emph{payload generator}, the \emph{payload corrector}, and the \emph{signature producer}.

The shadow model builder\circled{1} initiates the process by employing an open dataset containing both benign and malicious payloads to train a WAF shadow model designed to behave similarly to the WAF in question. The payload generator \circled{2} then leverages known malicious payloads, namely, those that can be detected by the WAF, to create new malicious payloads that may be capable of bypassing the shadow model's detection mechanism.

Subsequently, the payload corrector\circled{3} extracts payload keywords from the original dataset, removes any invalid payloads or modifies them to ensure their validity based on the distance between the extracted keywords\circled{4}. The payload corrector then forwards the valid payloads to both the shadow model\circled{5} and the real WAF\circled{6} to evaluate the generated payloads' effectiveness in evading detection.

The valid malicious payloads generated by the payload generator are initially tested against the shadow model. Those that successfully bypass the shadow model are then evaluated against the real WAF. Payloads that evade detection by both systems are utilized by the signature producer to build a payload signature. This approach ensures that the shadow model accurately reflects the real WAF's detection capabilities, and any discrepancies prompt updates to the shadow model for improved fidelity\circled{7}.
The signature producer combines the features of multiple valid signatures to create a single signature that is more robust and effective at detecting malicious payloads. Ultimately, the signature producer\circled{8} employs the generated signatures of malicious payloads to update the security rules of the real WAF, thereby enhancing its capability to detect and block potential threats.

In summary, \scheme provides a systematic and effective method for identifying and addressing previously unknown malicious payloads. By leveraging a combination of machine learning (ML) techniques and payload analysis, this approach offers a comprehensive and robust solution for improving the WAF security against the unknown mutated malicious payloads.

\subsection{Shadow Model Builder}
\label{sec:shadow}

We opted for a shadow model over direct WAF interaction due to several intertwined rationales. Firstly, most commercial WAFs operate as black-box systems, with their internal processes remaining proprietary, thus limiting direct adversarial sample generation. Secondly, continuous interactions with a live WAF can deplete its resources and hamper its efficiency. By using a shadow model, we conserve the WAF's resources, ensuring consistent performance. Lastly, the shadow model’s adaptability facilitates swift experimental adjustments and iterative testing, a feat challenging with static WAFs. Together, these reasons underscore the shadow model's efficiency and versatility in our study.

The objective of the shadow model builder is to train a deep learning shadow model that behaves similarly to the given WAF, i.e., the prediction results of the shadow model and the given WAF are similar. The WAFs can be viewed as a binary classifier, where the input payload (i.e., the sequence of characters) is classified as either benign or malicious. We use Convolutional Neural Networks (CNNs) as the model structure, which has demonstrated excellent performance in natural language processing and text classification~\cite{kim2014convolutional}. For instance, Shibahara et al.~\cite{shibahara2017malicious} proposed using CNNs to detect malicious URL sequences. Furthermore, CNNs outperform traditional non-deep learning methods in specific domains, due to their powerful feature representation, which does not require human expert knowledge and satisfies our requirements.

To build a valid CNN model, it is necessary to use an appropriate input data format. We use the payloads labeled by the original given WAF as the raw training data. However, attackers usually bypass detection by encoding the payloads with URL encoding, e.g., encoding \verb|<script>alert('XSS')</script>| as \verb|%3Cscript%3Ealert(%27XSS%27)%3C%2Fscript%3E|.
Using such payloads directly would reduce the accuracy of the WAF detection model.
We decoded malicious payloads before analysis to counter attackers' common use of URL encoding for evasion. This step, crucial for realism, ensures our model trains on data akin to what WAFs encounter, including decoded threats, thereby enhancing our findings' applicability and accuracy.
To better capture the true intent of these obfuscated payloads, we process all the payloads by decoding malicious payloads and keeping benign payloads unchanged.
Thus, the strategy ensures \scheme's robustness and relevance by effectively recognizing underlying attack patterns while maintaining practical applicability.
Our primary goal was not to artificially boost performance. Instead, we aimed to create a training set that mirrors real-world scenarios, wherein malicious and benign payloads exhibit discernible characteristics.
Then, we use Word2Vec~\cite{mikolov2013efficient} to learn a word embedding of the payloads. Word2Vec generates a vector in a low-dimensional embedding space, capturing semantic relationships between the corresponding words. The payloads are taken as input for Word2Vec, and the resulting vectors are used as input data for the CNN model.

In web security, using shadow model provides a strategic advantage when enhancing WAFs. A shadow model is designed to emulate the target WAF's response to specific attack types, using similar data for training to replicate its behavior accurately. This approach provides a focused and non-intrusive testing environment, allowing for in-depth analysis without compromising the operational integrity of the actual WAF.
This method allows for detailed examination without revealing the WAF's vulnerabilities or configurations. Through automated refinement, shadow models connect research with real-world use, keeping the WAF secure and private while addressing and mitigating potential security oversights.

The CNN model consists of four layers: convolution, max-pooling, fully-connected, and Softmax. The convolution layer applies filters to the input matrix, while the max-pooling layer reduces the dimensionality of each feature map, retaining essential information. The fully-connected layer extracts high-order features from the input, and the Softmax layer outputs a label (0 or 1) to determine if the payload is benign or malicious.

Our CNN architecture was chosen for its simplicity and suitability for our payload data. While deeper architectures capture hierarchical features well in complex data like images, our simpler model effectively captures the necessary patterns in the payloads, ensuring robust performance without overfitting. Preliminary experiments with deeper networks did not significantly improve performance and increased computational costs. Input size of CNN was set to 256 units in FC layer, and embedding vector dimension of 128 were found to be optimal. A dropout layer with a probability of 0.3 was added after pooling to prevent overfitting. The shadow model was trained using binary cross-entropy loss, with Adam as the optimizer, a learning rate of 0.05, and 50 training epochs.

\subsection{Payload Generator}
\label{sec:payload}
The goal is to generate adversarial examples against the shadow model, which will be used to test the real WAF. The payload generator takes in the payloads labeled as malicious by the original WAF and outputs the adversarial examples that can bypass the shadow model.
The payloads that fool the shadow model might also fool the original WAF.

Since the maliciousness of a payload depends on the context, i.e., the preceding and succeeding payloads, we use Recurrent Neural Networks (RNNs) as the payload generator. RNNs are used to model sequential data, such as for text generation, language modeling, and music prediction, and have indeed been used for program analysis~\cite{gu2016deep,Li2018}. The ability to learn sequential structures, where dependencies to preceding inputs exist, is an important characteristic when learning input format structures for generation. Compared to traditional n-gram-based approaches~\cite{cavnar1994n} that are limited by contexts of finite length, RNN allows for learning arbitrary length contexts to predict the next sequence of characters. Given a corpus of malicious payloads, the RNN model can be trained in an unsupervised manner to learn a generative model to produce new malicious payloads.

RNN first requires that categorical values be mapped to integer values, where every unique word or token from the payload undergoes an assignment to a unique integer identifier. The token-to-integer mapping prepares the textual data for subsequent neural network processing.
Let X = {$x_1$,$x_2$,...,$x_N$} be the input sequence with $x_t \in \mathbb{N} \mid 1 \leq t \leq N$, where t represents the position in the input sequence. The input module takes $x_t$ and transforms it into a one-hot coded vector $\hat{x}_t \in \mathbb{N}^I$ with I = max(X). Each integer value is represented as a binary vector that has all zero values except the index of the integer, which is marked with 1. Converting categorical data, namely $x_t$, is necessary, and those categories are handled as features during the training process.

The recurrent layers use Gated Recurrent Unit (GRU) nodes~\cite{cho2014learning} to solve the vanishing or exploding gradient problem that RNNs suffer from. At each step, the RNN model accepts a vector (i.e., tokenized sequences) as input.
The input, as well as the previous state of the hidden layer, is used to predict the output. The result provides the probability distribution for predicting the next value of the input sequence.

We use cross-entropy as the loss function, and the model is trained to minimize the loss. Adam~\cite{kingma2014adam,wu2020liveness} is used as the optimizer. The learning rate is initially set to 0.001, and it is reduced to half every 10 training epochs. The batch size is 512, and the GRU units are set to 256. We train each model for 50 epochs, and the dropout probability is 0.3.

To validate the malicious nature of our derived payloads, we employed a three-pronged approach: Firstly, static analysis using established detection tools confirms the presence of malicious signatures post-mutation. Secondly, behavioral analysis in controlled environments ensures payloads' harmful potential and identifies benign transformations. Lastly, comparative analysis against original counterparts verifies consistency in malicious behavior. This rigorous validation process underscores the reliability of our findings and our commitment to enhancing WAF effectiveness.

To mutate a malicious payload using RNN, we can provide an initial substring, such as \texttt{select}, and the RNN model can predict and generate the following substring, enabling us to achieve the goal of mutating the malicious payload. As an example of an SQL injection attack, given the keyword \texttt{select}, the RNN model can generate both valid and invalid injection payloads.

\begin{tcolorbox}
	\scriptsize
	\textbf{Example of generated payload 1}:\\
	\texttt{select * from users where username = “abc”}  (valid) \\
	\textbf{Example of generated payload 2}: \\
	\texttt{select * fram text} (invalid)
\end{tcolorbox}

However, some generated payloads may be invalid and therefore inexecutable. These payloads are passed to the payload corrector to either make them valid or remove them.

\subsection{Payload Corrector}
\label{sec:corr}

The goal of the payload corrector is to correct or exclude the invalid payloads generated by the payload generator.
Basically, the payload corrector modifies keywords to ensure that the generated malicious payloads can evade the detection of the WAF.
It finds the high-frequency keywords from the original datasets.
Finally, it modifies the generated payloads based on the keywords or excludes the payload to ensure that the generated payload can bypass the shadow model.

\textbf{Edit distance measuring.}  We observed that specific keywords, such as \texttt{SELECT}, \texttt{FROM}, frequently appear across various payloads. These keywords often form the backbone of the malicious intent in many attacks.
Therefore, we prioritized these recurring keywords for our analysis.
We count the frequency for each substring and select the substrings with higher frequency as the keywords.
Specifically, we select five keywords, since these five represented the most commonly recurring and impactful terms in our dataset, while other keywords show a quite small frequency comparing with the five keywords.
To determine whether the substrings in the payload needs to be modified, we measure the \emph{edit distance} between the keywords and substrings in the generated payloads.
Edit distance is a substring metric to quantify how dissimilar two substrings (e.g., words) are.
It is measured by counting the minimum number of operations (e.g., insertion, Deletion, and replacing) required to transform one substring into the other.
Specifically we use Levenshtein distance as the distance metric~\cite{schulz2002fast}.
The payload is first split into substrings with the whitespace as separator.
Then we calculate the Levenshtein distance between the keywords and each split substrings as the correcting value.


\textbf{Payload correcting.}  The measured edit distance is compared with the predefined thresholds to remove invalid payloads.
We also set two thresholds for two different conditions, i.e., a lower threshold for correcting the invalid payloads and an upper threshold for excluding the invalid payloads.
If each substring in payloads has a higher edit distance than the upper threshold,
we exclude the payloads and never use them for our subsequent testing because these payloads are far from the effective payload, and they may be invalid even though we correct these payloads.
For the payloads that have substrings with a correcting value higher than the lower threshold,
we modify the payloads and reduce the edit distance by deleting, inserting, swapping, or replacing them until they are lower than the lower threshold.
Because these payloads may be effective after correcting, we do not change the payloads in which all the substrings have lower edit distance, i.e., these payloads can be regarded as effective.
We also exclude the short payloads.

The output payloads are then input to the shadow model to find the adversarial examples.
The algorithm of the payload corrector is shown as Algorithm~\ref{alg:1}.
Specifically, the lower and upper threshold are 4 and 8.
We assume that the payloads that deceive the shadow model can also deceive the WAF,
which is consistent with common sense and has been verified in our experiments.

In WAFBooster, the payload generator can craft adversarial examples with intentional misspellings or obfuscations, such as 'SALECT,' to simulate evasion tactics used by attackers. This method tests the WAF's ability to detect and block both direct and subtly altered attack vectors.
As an example of SQL injection payload \texttt{salect * from users where username = "abc"},
by using whitespace as the separator,
the payload can be split as eight substrings,
including \texttt{salect}, \texttt{*}, \texttt{from}, \texttt{users}, \texttt{where}, \texttt{username}, \texttt{=}, and \texttt{"abc"}.
Then we calculate the Levenshtein distance between keywords and split tokens,
where the Levenshtein distance between the token \texttt{salect} and \texttt{select} is 1.
While for the token \texttt{from} and \texttt{where},
the Levenshtein distance of 0 is 0.
Thus, the Levenshtein distance of this payload is 1.

Afterward, the payload corrector outputs the valid malicious payloads that can bypass the shadow model and real WAF.
These payloads will be used to produce the signature.

\begin{algorithm}[!t]
	\setstretch{.95	}
	\scriptsize
	\caption{Payload Corrector}
	\label{alg:1}
	\begin{algorithmic}
		\REQUIRE
		Payload: $x$.
		Upper threshold: $UT$.
		Lower threshold: $LT$.
		Top tokens, $tokens$.
		\ENSURE
		Output  payload $x^{*}$

		\STATE $substringlist \gets \textit{split}(x)$

		\WHILE{$substringlist != null$}
		\STATE $t \gets \textit{Levenshtein}(substringlist,tokens)$

		\IF{$t > UT$}
		\STATE $\textit{Discard}(substringlist)$
		\ELSE
		\WHILE{$t > LT$}
		\STATE $keyword \gets \textit{Correct}(substringlist)$
		\ENDWHILE
		\STATE $x^{*} \gets \textit{Replace}(x,substringlist)$
		\ENDIF
		\ENDWHILE
	\end{algorithmic}
\end{algorithm}

\subsection{Signature Producer}
\label{sec:signature}

Traditional signatures are carefully designed manually by security experts.
The goal of signature producer is to automatically produce the malicious signature that matches the mutated evasion attacks for WAF.
Signatures are widely used to protect from attacks with a basic assumption:
there exists a single payload substring,
which remains unchanged in the process and unique enough to be used as a signature without causing false rejections.
This module takes in the bypassing malicious payloads against WAF, and generate rules to update the security policies for the given WAF.

\textbf{Finding important substrings.}
Considering a large search space,
we aim to find the substrings showing high importance on the WAF decision results.
Algorithm~\ref{alg:2} present the algorithm of signature producer.
Specifically, we employ four scoring functions to measure the importance.
The scoring function needs to have the two characteristics:
(1) it can correctly reflect the importance of the substring to the prediction, (2) the computation should be with high efficiency.

We assume the input sequence x = $x_1x_2...x_n$,
where $x_i$ is the $i_{th}$ token.
We design scoring functions to evaluate the impact of the token in the payload on WAF decision results.
Specifically, we consider four scoring function: delete, replace, head, and tail score function.

\begin{algorithm}[!t]
	\setstretch{.95}
	\scriptsize
	\caption{Find Important Token}
	\label{alg:2}
	\begin{algorithmic}[1]
		\REQUIRE
		The input payload $x = x_1x_2...x_n$
		\ENSURE
		Important tokens $x_i^{*}$
		\STATE $x_i \gets \textit{Select}(x)$
		\WHILE {$x_i != null$}
		\STATE $DS(x_i) \gets f(x_1...x_{i-1}x_ix_{i+1}...x_n)-f(x_1...x_{i-1}x_{i+1}...x_n)$
		\STATE $RS(x_i) \gets  f(x_1...x_{i-1}x_ix_{i+1}...x_n)-f(x_1...x_{i-1}x'_ix_{i+1}...x_n)$
		\STATE $HS(x_i) \gets  f(x_1x_2...x_{i-1}x_i) - f(x_1x_2...x_{i-1})$
		\STATE $TS(x_i) \gets  f(x_ix_{i+1},...x_n) - f(x_{i+1},...x_n)	$
		\STATE $Score[i] \gets \textit{Add}(DS(x_i),RS(x_i),HS(x_i),TS(x_i))$
		\STATE $x_i \gets \textit{Select}(x)$
		\ENDWHILE
		\STATE $x_i^{*} \gets \textit{FindTopMToken}(Score[])$
	\end{algorithmic}
\end{algorithm}

\begin{figure}[!t]
	\centering
	\includegraphics[width=.9\linewidth]{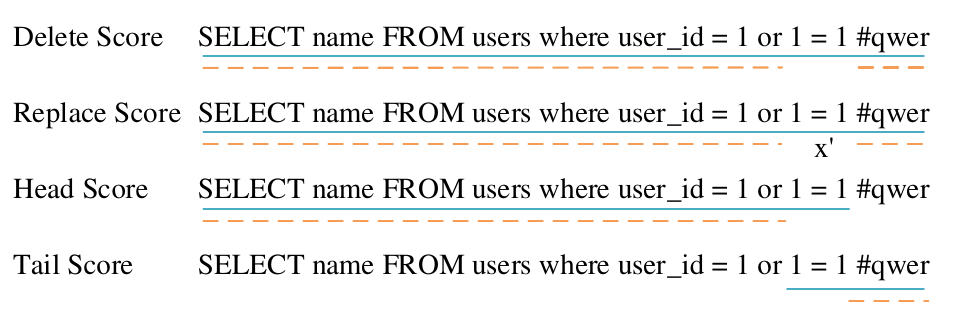}
	\caption{Example of scoring the token \texttt{1=1} in \texttt{SELECT name FROM users WHERE user\_id=1 or 1=1 \#qwer}. The WAF result is 1 if the blue and red parts differ, and 0 if they are the same.}
	\label{Fig.scoringFunction}
	\vspace{-5mm}
\end{figure}

\begin{inditemize}
	\item \textbf{Delete scoring function.}
	We compare the payload before and after deleting the token to evaluate the impact of token $x_i$ on WAF.
	Specifically, we compare the decision results based on the payload before and after deleting token i, e.g., $x_1x_2...x_i...x_n$ and $x_1x_2...x_{i-1}x_{i+1}...x_n$.
	The delete scoring formulation is given as Eq.~\ref{eq:ds}.
	\begin{equation}
		\begin{aligned}
			DS(x_i) = f(x_1x_2...x_i...x_n) - \\f(x_1x_2...x_{i-1}x_{i+1}...x_n)
			\label{eq:ds}
		\end{aligned}
	\end{equation}
	where \emph{f}($\bullet$) represents the discrimination result of WAF.

	\item  \textbf{Replace scoring function.}
	We replace the certain token to measure its impact.
	Specifically, we compare the decision results based on the payload before and after the replacement, e.g., $x_1x_2,...x_i...x_n$ and $x_1x_2...x'_i...x_n$,
	where $x'_i$ is the replaced token and we fix it as `unknown' to simplify the computation.
	The replace scoring formulation is given as Eq.~\ref{eq:rs}.
	\begin{equation}
		\begin{aligned}
			RS(x_i) = f(x_1x_2,...x_i...x_n) - \\f(x_1x_2...x'_i...x_n)
			\label{eq:rs}
		\end{aligned}
	\end{equation}

	\item \textbf{Head scoring function.}
	Since the tokens in the payload are sequential, the order in which the tokens appear is relevant and meaningful.
	We assess the significance of different sequences of header tokens by means of a head scoring function. This function compares the decision outcomes of header token sequences between $x_1x_2...x_{i-1}x_i$ and $x_1x_2,...,x_{i-1}$. The head scoring function is formulated as shown in Eq.~\ref{eq:hs}.

	\begin{equation}
		HS(x_i) = f(x_1x_2...x_{i-1}x_i) - f(x_1x_2,...,x_{i-1})
		\label{eq:hs}
	\end{equation}

	\item  \textbf{Tail scoring function.}
	Similar to head scoring function, the tail function compares the decision results of different tail tokens sequences.
	Specifically, it compares the tail tokens sequences between $x_ix_{i+1},...x_n$ and $x_{i+1}...x_n$.
	The formulations of tail scoring function is given as Eq.~\ref{eq:ts}.

	\begin{equation}
		TS(x_i) = f(x_ix_{i+1},...x_n) - f(x_{i+1}...x_n)
		\label{eq:ts}
	\end{equation}
\end{inditemize}

The results of the four functions can be represented by 0 and 1,
which means whether the discriminant results are affected when the token changes.
0 means no impact on WAF discrimination, otherwise 1 means it has an impact.
We comprehensively add on the results of the four scoring functions, evaluate all tokens, select several substrings that appear frequently and have a greater impact on discrimination, then use regular expressions to generate signatures. Each signature is made up of one or more such tokens.

As an example of SQL injection payload in Fig.~\ref{Fig.scoringFunction},
it illustrates the process of evaluating the importance of \texttt{1=1} in the payload
\texttt{SELECT name FROM users WHERE user\_id=1 or 1=1 \#qwer}.
The solid and dashed underlines are the tokens by the scoring function, respectively.
For example, to evaluate the token importance of  \texttt{1=1},
the delete scoring function compares discrimination result between \texttt{SELECT name FROM users WHERE user\_id=1 or \#qwer} and \texttt{SELECT name FROM users WHERE user\_id=1 or 1=1 \#qwer}.
The tail scoring function compares discrimination result between
\texttt{\#qwer}
and \texttt{1=1 \#qwer}.

Next step is to generate signatures. It is believed regular expression has significant advantages for attack detection in terms of flexibility, accuracy, and efficiency. However, the full regular expression is too complex and its numerous syntax rules are not needed for signature. Hence, we introduce the simplified regular expression as a way of representing signatures. A simplified regular expression signature is a simplified form of a regular expression that contains only one qualifier, "$\backslash$S*", which represents any non-white-space character.

\textbf{Signature matching.}
Edit distance is used to cluster these tokens and assign similar tokens to different groups.
Then we find the longest subsequence that is common to all the given tokens in the same group,
where the common subsequence should be with the same order.
It does not require consecutive, but matches as many consecutive substrings as possible.

We regard the regular expression with a higher matching score as the common subsequence.
To fulfill this goal, we use the  Smith-Waterman algorithm~\cite{ligowski2009efficient}
and define the matching score to measure the matching quality.
Specifically, we add one to each matching character while subtracting gap penalty $g_p$ if there are one non-matching character, where $g_p$ is set as 0.8 in our work.
The algorithm rewards matching and penalizes insertion or deletion, with the matching score set to 1 and the insertion or deletion penalty set to 0.8 in our case.

As an example, for the substrings `aaselectaaafromaaa' and `selectbbbbfrombbb', the longest common subsequence is `select from',
where the regular expression is \texttt{\textbackslash S*select\textbackslash S*from\textbackslash S*}.
Besides, `saealaeacatafroma' and `bfrombsbeblbebcbt' can match out two different regular expression,
including \texttt{\textbackslash S*s\textbackslash S*e\textbackslash S*l\textbackslash S*e\textbackslash S*c\textbackslash S*t\textbackslash S* } and
\texttt{\textbackslash S*from\textbackslash S*}.
The former regular expression is with $6-5*0.8=2$, and the later one is with $4$.
Thus, we select the later one as the continuously matched signatures.

We iteratively perform sequence matching for different groups and find all matching sequence.
To output the malicious signature, we then add \texttt{\textbackslash S} between every two matching sequence to generate a simplified regular expression.

\section{Performace Evaluation}
\label{sec:Performace}
This section reports the performance of \scheme.

\subsection{Experimental Setup}
\textbf{Real WAF selection.}
To enhance the WAF security, we first selected real WAFs that fulfill these goals as targets.
1) it should come from the commercial company or be a highly rated project on GitHub~\cite{githubwaf} with more than 500 stars
that reports no errors during running.
2) it can disable the protection of challenge collapsar attack, since this protection prevents our system from building simulator and testing the WAF efficiently.
We finally selected the following eight WAFs, and performed comprehensive experiments on these real WAFs.

\emph{ModSecurity}~\cite{ModSecurity},  an open source, cross-platform WAF engine for Apache, IIS and Nginx,
is one of the most widely deployed WAFs available.
It allows to use the security rule language to monitors, logs and filters HTTP traffic in the runtime.
\emph{Naxsi}~\cite{Naxsi}, an open source third party Nginx module, and is available for UNIX-like platforms.
\emph{Lua-resty-waf}~\cite{lua-resty-waf} uses Nginx Lua to analyze HTTP requests and process against a flexible rule structure.
\emph{Safedog}~\cite{Safedog} provides real-time website security protection and covers a variety of security features,
including active security protection, resource protection. It is maintained by a cloud security company.
\emph{Xwaf}~\cite{xWAF} is a cloud WAF for medium and small businesses that supports defense against common WEB attacks.
\emph{Openwaf}~\cite{openwaf}, the first fully open source WAF based on \texttt{nginx\_lua API}.
\emph{Verynginx}~\cite{VeryNginx}  is a powerful and friendly Nginx based on \texttt{openrestry}.
\emph{Unixhot}~\cite{unixhot} is an open source third-party Nginx module, and available for UNIX-like platforms.

\textbf{Dataset - benign payloads.}
We used the benign data from a public dataset, HTTP dataset CSIC 2010~\cite{dataset}.
It is generated automatically and contains 36,000 normal requests targeted to an e-Commerce web application.
Besides, we generated benign queries on our servers to collect benign payloads as the benign dataset.
We used the combination of the public dataset and generated benign datasets as benign payloads for performance evaluation.
The collected benign dataset was used for three attacks.

\textbf{Dataset - malicious payloads.}
Due to privacy issues and the commercial value of malicious datasets maintained by the WAF security company,
there are no available and proper malicious payloads in our tasks~\cite{IDS2018}.
To generate malicious payloads using the benign dataset,
we used the generation grammar~\cite{demetrio2020waf} that supports a variety of SQL queries.
For each terminal symbol (i.e., tables \texttt{t}, fields \texttt{f}, values \texttt{v}),
we used the values from different dictionaries,
which are possible values of the columns in the database.
The grammar is given as follows:
\begin{tcolorbox}
	\scriptsize
	1. \texttt{\textit{Q} ::= \textit{S} | \textit{U} | \textit{D} | \textit{I}}

	2. \texttt{\textit{S} ::= \textbf{SELECT} ( $\bar{f}$ | *) \textbf{FROM} \textit{t} \textbf{WHERE} \textit{e} [\textbf{LIMIT} $\bar{v}$]}

	3. \texttt{\textit{U} ::= \textbf{UPDATE} \textit{t} \textbf{SET} \textit{f}=\textit{v} \textbf{WHERE} \textit{e} [\textbf{LIMIT} $\bar{v}$]}

	4. \texttt{\textit{D} ::= \textbf{DELETE} \textbf{FROM} \textit{t} \textbf{WHERE} \textit{e} [\textbf{LIMIT} $\bar{v}$]}

	5. \texttt{\textit{I} ::= \textbf{INSERT} \textbf{INTO} \textit{t}($\bar{f}$)\textbf{VALUES} ($\bar{v}$)}

	6. \texttt{\textit{e} ::=  \textit{f}  $\gtreqless$ \textit{v} | \textit{f} \textbf{LIKE} \textit{s} | \textit{e} \textbf{AND} \textit{e'} | \textit{e} \textbf{OR} \textit{e'}}
\end{tcolorbox}
The query \emph{Q} contains \emph{select} \texttt{S}, \emph{update} \texttt{U}, \emph{delete} \texttt{D},
or \emph{insert} \texttt{I}.
\texttt{S}, \texttt{U}, and \texttt{D} can end with \texttt{LIMIT} optionally.
These queries are with several types of parameters, including fields \texttt{f}, tables \texttt{t}, values \texttt{v},
substrings \texttt{s} and boolean expressions \texttt{e} and \texttt{e'}.
$\bar{\cdot}$ denotes a finite vector with comma separated.
The values of \emph{t} and \emph{f} are taken from a real target web application database.
For \emph{v} and \emph{s}, we used different values depending on the type of the generated query.
With these grammars, we used the scripts extracted from penetration testing tools to generate malicious payloads.

For SQL injection, we used random query generator~\cite{randgen},
a testing tool for SQL Server to generate payloads and run the scripts to make it malicious.
With the given grammar,
it returns a set of payloads that belong to the language denoted by the grammar.
We also extract scripts from existing penetration testing tools, including sqlmap~\cite{sqlmap}
and OWASP ZAP~\cite{OWASPTen}, to generate malicious payloads.
For the command injection attack, we modified Commix~\cite{Commix} to generate malicious payloads for evaluation.
It is an open-source penetration testing tool that automates the detection and exploitation of command injection vulnerabilities.
For XSS attacks, we used XSStrike~\cite{XSStrike} to perform similar operations.
XSStrike is an advanced open-source cross-site scripting detection suite.

Our dataset is a diverse amalgamation of web request payloads from multiple sources, including real-world web applications, traffic logs, and public datasets. The data was split into a 70\% training set, 15\% validation set, and 15\% testing set, ensuring stratified sampling for even representation. The dataset's composition consists of 40\% benign and 60\% malicious payloads, with varied attack vectors like SQL injection and XSS. Preprocessing steps, including tokenization, normalization, and vectorization, were employed, while synthetic data generation and augmentation strategies ensured balanced representation. We've also integrated external datasets for auxiliary validation to enhance our model's robustness.

\begin{table}[!t]
	\renewcommand\arraystretch{0.95}
	\centering
	\scriptsize
	\caption{Performance of Shadow model and \# corrected examples (CE) by payload corrector in \scheme on eight typical WAFs under SQL injection (SQLi), cross-site scripting (XSS), and command injection (CI) attacks.
	}
	\begin{tabular}{lcccccr}
		\hline
		                                         &      & \multicolumn{2}{c}{{\textbf{Original payloads}}} & \multicolumn{2}{c}{{\textbf{Shadow model}}} &                        \\ \cline{3-6}
		\multirow{-2}{*}{{\textbf{WAF}}}         &
		\multirow{-2}{*}{{\textbf{Attack Type}}} &
		{\textbf{TRR}}                           &
		{\textbf{TAR}}                           &
		{\textbf{TRR}}                           &
		{\textbf{TAR}}                           &
		\multirow{-2}{*}{{ \#  \textbf{CE}}}                                                                                                                                      \\
		\hline
		                                         & SQLi & 82.9\%                                           &                                             & 97.7\% & 85.2\% & 4033 \\
		                                         & XSS  & 95.8\%                                           &                                             & 98.1\% & 97.2\% & 4959 \\
		\multirow{-3}{*}{ModSecurity}            & CI   & 84.0\%                                           & \multirow{-3}{*}{98.4\%}                    & 98.4\% & 89.6\% & 3289 \\
		\hline
		                                         & SQLi & 94.8\%                                           &                                             & 99.3\% & 98.7\% & 4808 \\
		                                         & XSS  & 95.4\%                                           &                                             & 98.7\% & 95.3\% & 3437 \\
		\multirow{-3}{*}{naxsi}                  & CI   & 72.3\%                                           & \multirow{-3}{*}{67.6\%}                    & 87.8\% & 93.8\% & 2274 \\
		\hline
		                                         & SQLi & 100\%                                            &                                             & 91.2\% & 93.1\% & 3397 \\
		                                         & XSS  & 98.9\%                                           &                                             & 97.6\% & 97.2\% & 2348 \\
		\multirow{-3}{*}{lua-resty-waf}          & CI   & 82.2\%                                           & \multirow{-3}{*}{67.5\%}                    & 96.4\% & 97.8\% & 2991 \\
		\hline
		                                         & SQLi & 88.4\%                                           &                                             & 97.2\% & 83.8\% & 4327 \\
		\multirow{-2}{*}{safedog}                & XSS  & 92.5\%                                           & \multirow{-2}{*}{97.6\%}                    & 95.3\% & 96.0\% & 2549 \\
		\hline
		                                         & SQLi & 76.6\%                                           &                                             & 92.4\% & 90.1\% & 3884 \\
		\multirow{-2}{*}{xwaf}                   & XSS  & 94.8\%                                           & \multirow{-2}{*}{81.3\%}                    & 99.3\% & 96.4\% & 2474 \\
		\hline
		                                         & SQLi & 72.9\%                                           &                                             & 94.9\% & 84.4\% & 3845 \\

		\multirow{-2}{*}{openwaf}                & XSS  & 89.9\%                                           & \multirow{-2}{*}{99.8\%}                    & 99.2\% & 95.3\% & 3105 \\
		\hline
		verynginx                                & SQLi & 45\%                                             & 81.4\%                                      & 99.7\% & 95.0\% & 3634 \\
		\hline
		unixhot                                  & XSS  & 92.9\%                                           & 99.9\%                                      & 99.1\% & 94.2\% & 2451 \\ \hline
	\end{tabular}
	\label{tab-Evaluation}
\end{table}

\textbf{Evaluation metric.}
If the malicious payloads bypass the WAF, we think the WAF is vulnerable to attacks.
We define the ratio of the number of truly rejected payloads and the number of all malicious payloads as true rejection rate (TRR),
which measures the security against malicious payloads.
The ratio of the number of truly accepted payloads and the number of all benign samples is true acceptance rate (TAR).
The ratio of the number of falsely accepted payloads and the number of all malicious payloads is false acceptance rate (FAR).
The ratio of the number of falsely rejected payloads and the number of all benign payloads is false rejection rate (FRR).

Experiments were conducted on a Linux workstation with an Intel Xeon CPU E5-2680 with 32GB RAM.
Selected real WAFs were deployed on the Linux workstation or another Windows workstation with an same hardware configuration.
The model was implemented using TensorFlow.

\subsection{Overall Performance}
\label{sub:detect}

Table \ref{tab-Evaluation} reports TRR and TAR of \scheme on eight typical WAFs under SQL injection (SQLi), cross-site scripting (XSS), and
command injection (CI) attacks.
Original payloads refers to the input data that has not undergone any modification or preprocessing before being fed to the WAFs or the shadow model for detection.
Since some WAFs cannot resist certain attacks, we did not test all eight WAFs against
SQL injection, cross-site scripting, and command injection attacks.
As shown in the table, only ModSecurity, naxsi, and lua-resty-waf can resist command injection attacks.
The possible reason is that command injection is required to be combined with other attacks and is relatively rare in web applications.
Some WAFs, e.g., naxsi, have strict rules which reject not only malicious attacks but also benign requests.
So we modify the parameters to make its blocking capability reach a reasonable level for testing.

\textbf{Performance of the original payloads.}
Table~\ref{tab-Evaluation} also presents the TRR of malicious payloads and TAR of benign payloads under the original payloads.
The results show that most WAFs are with a high TRR against malicious payloads.
For example, openwaf is with a TRR of 72.9\% against SQL injection and 89.9\% against XSS.
Note that verynginx is with a low TRR of 45\% against SQL injection.
We find that verynginx only provides basic protection rules and relies on manually adding protection rules.
We can also see that the TAR of most WAFs is high.
It can be found from the payloads that not all passing rates have reached a high level because most WAFs choose to sacrifice FRR
for the blocking rate.
For the blocked payloads and the bypassing payloads,
we select part of the payloads to train the deep learning model and keep the balance of the training set.

\textbf{Performance of the shadow model.}
We used the validation payloads from the original validation dataset to evaluate the performance of the shadow model.
The column of `shadow model' in Table \ref{tab-Evaluation} presents the TRR and TAR of the shadow model.
The shadow models achieve more than 80\% of TRR and TAR, especially most of them reach above 95\%.
As an example of ModSecurity, it achieves a TRR of 97.7\% and a TAR of 85.2\% under SQL injection.
Generally, the shadow model outperforms the original WAF or at least achieves a similar performance.

\textbf{Effectiveness of the payload corrector.}
Since the GRU model generates invalid payloads for our testing, which may deviate far from the grammatical structure, we exclude or modify the generated payloads.
The column of  `\# corrected examples' in Table \ref{tab-Evaluation} indicates the number of adversarial examples we generated after correction.
It can be seen that the fusing of payload generator and payload corrector in \scheme can generate more than an average of 2000 malicious payloads for each WAF.
As an example of ModSecurity, it generates 4033, 4959, and 3289 samples for SQL injection, XSS, and command injection attacks.
It can be seen that the number of SQL injection attacks is more than the other two kinds of attacks after correction.
We speculated that the payload of SQL injection attacks is more structured.

\begin{figure*}[!t]
	\centering
	\subfloat[SQL injection attack]{\includegraphics[width = 0.29\linewidth]{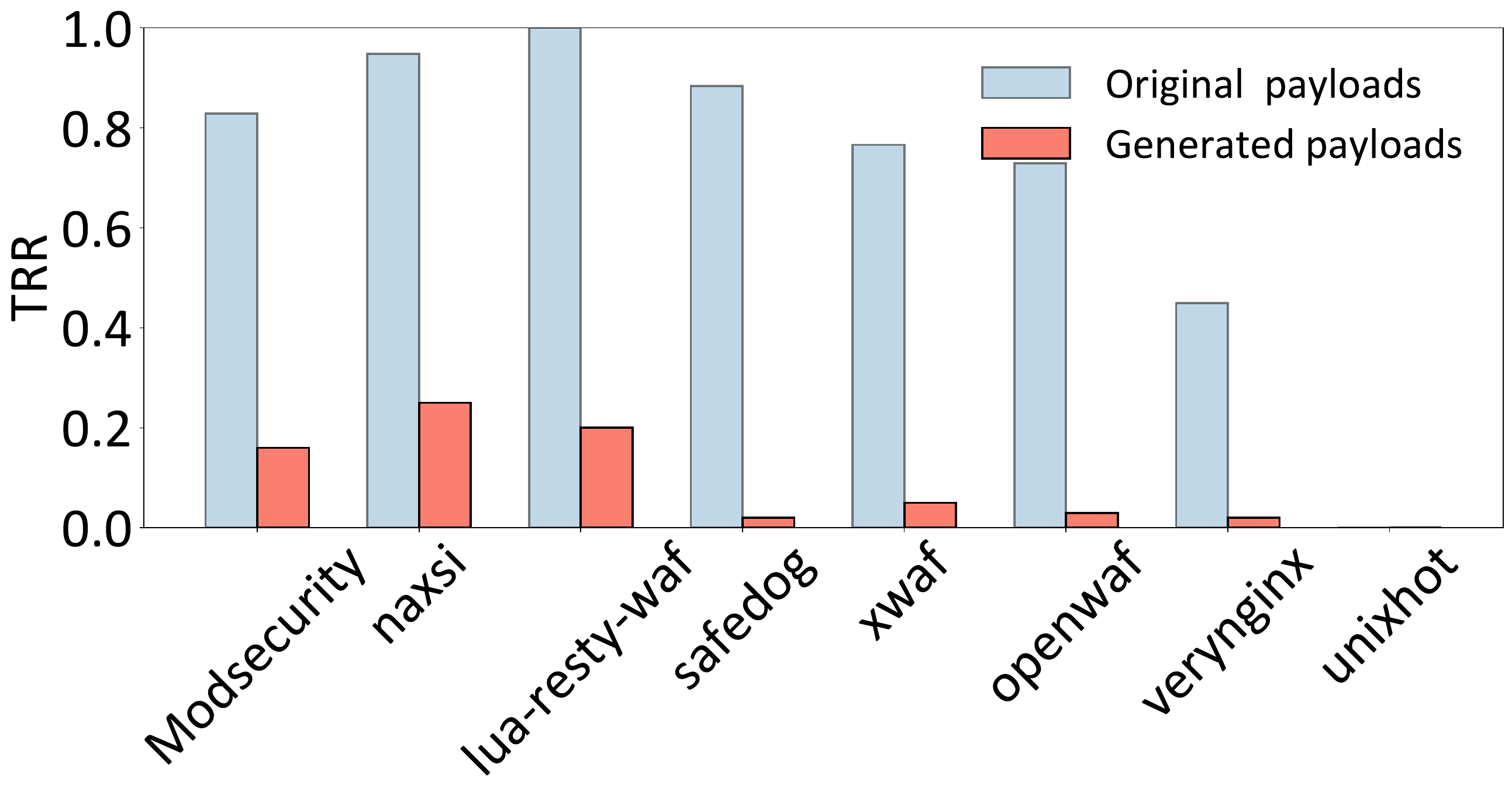}} \hspace{3mm}
	\subfloat[Cross-site scripting attack]{\includegraphics[width = 0.29\linewidth]{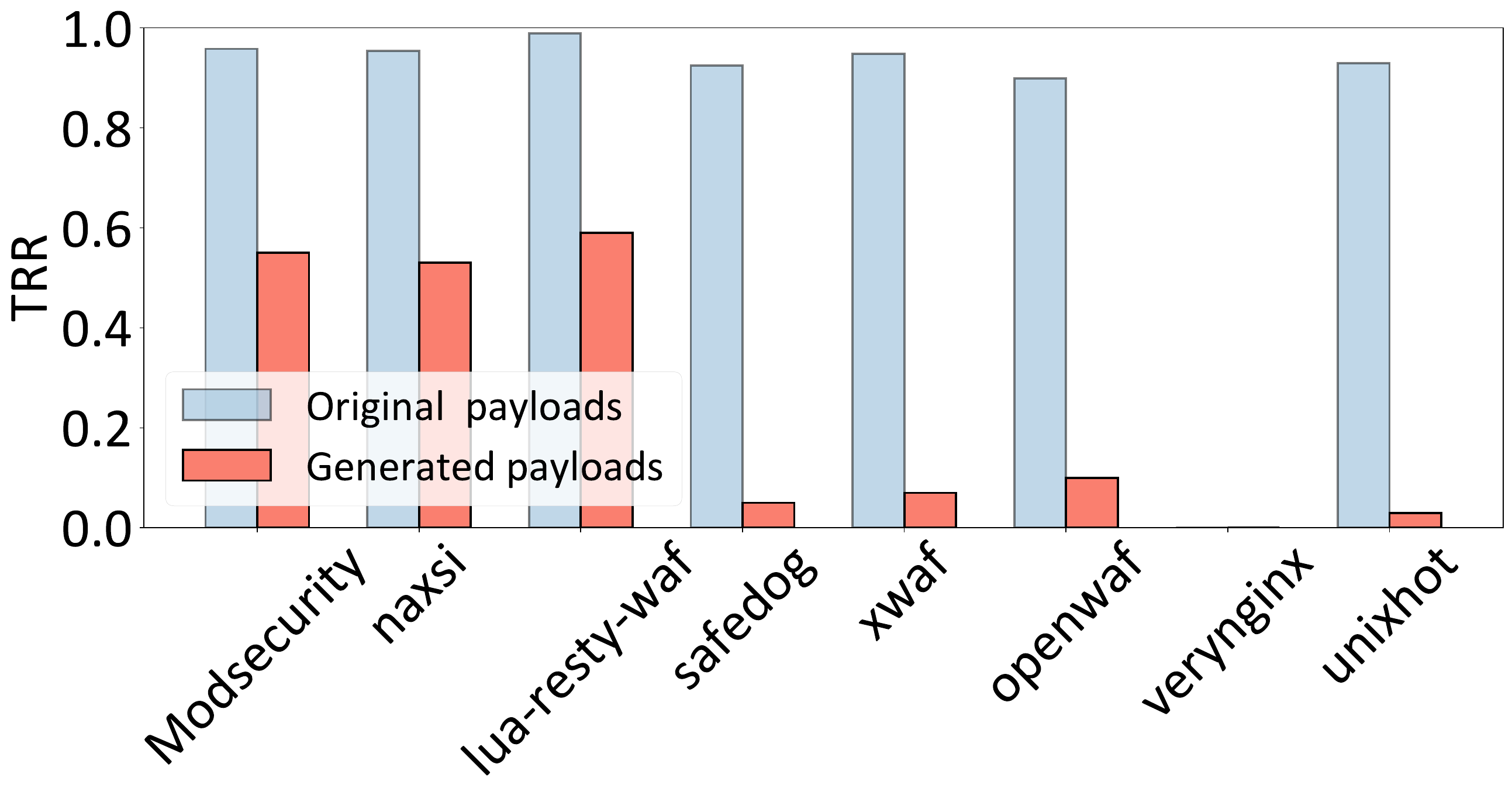}}\hspace{3mm}
	\subfloat[Command injection attack]{\includegraphics[width = 0.29\linewidth]{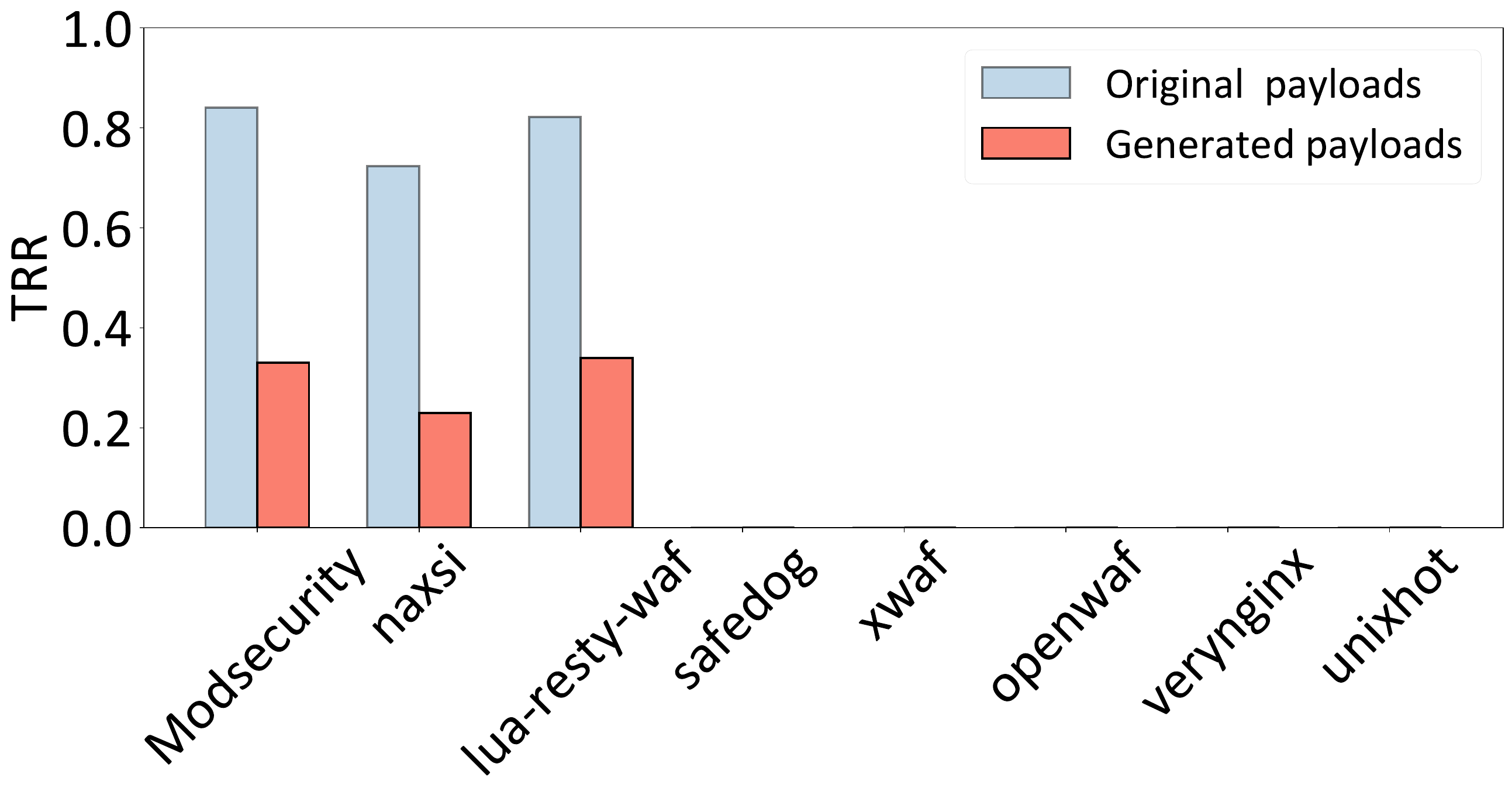}}
	\caption{TRR of original and generated payloads against  SQL injection, cross-site scripting, and command injection attacks}
	\label{fig:tr_payloads}
\end{figure*}

\textbf{Performance of generated malicious payloads.}
We evaluated the ratio of the number of truly rejected samples by the real WAF and the number of \emph{all generated malicious payloads} as TRR.
We used the corrected malicious payloads for evaluation.
Fig.~\ref{fig:tr_payloads} shows the comparison of the TRR of original (baseline) and generated payloads against  SQL injection,
cross-site scripting, and command injection attacks on eight practical WAFs.
The results show that the TRR of generated malicious payloads against all three kinds of attacks is significantly smaller than the original payloads.
For example, the TRR of unixhot against malicious XSS payloads is 3\%, while its TRR of against original malicious payloads is 92.9\%.
For example, the TRR of safedog against malicious XSS payloads is 2\%, while its TRR against original malicious payloads is 88.4\%,
The results demonstrate the effectiveness of our generated malicious payloads against real WAFs.

\subsection{Generated Malicious Payload Cases}
\label{sub:cases}

Determining if a payload is malicious depends on the specific web application, making it challenging to confirm if all bypassing attack payloads are usable in real-world scenarios. However, we double-checked the bypassing payloads identified by our system. We examined their effectiveness in certain situations and assessed our system's contribution. Some bypassing attacks did pose a threat to web applications. Below are examples of bypassing attacks, particularly against ModSecurity and Safedog, which are widely used WAFs.

\textbf{Generated SQL injection payload case.}
Code 1(a) shows a malicious SQL injection payload that can bypass the detection of Safedog.
In this case, Safedog can block substrings similar to the keywords to prevent case swapping.
However, we found that the malicious payload uses \texttt{/*!  */} inline comments and the substitution of \texttt{\%0A} for the space character.
The web application can properly identify inline comments and the substitution for the space character, but WAF regards it as a benign query,
which eventually led to the successful bypass of the WAF.

\begin{figure*}[!t]
	\centering
	\stepcounter{codefigure}
	\captionsetup{name=Code, labelsep=colon} 
	\subfloat[SQLi case]{\includegraphics[width = 0.27\linewidth]{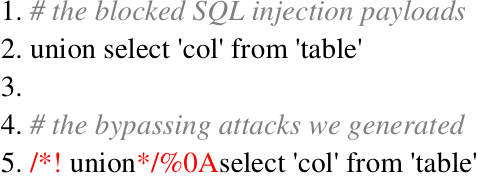}} \hspace{4mm}
	\subfloat[XSS case]{\includegraphics[width = 0.27\linewidth]{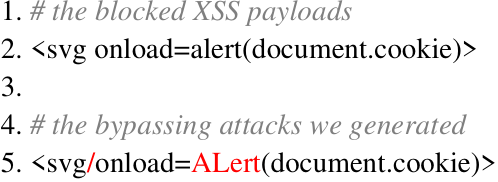}}\hspace{4mm}
	\subfloat[Command injection case]{\includegraphics[width = 0.27\linewidth]{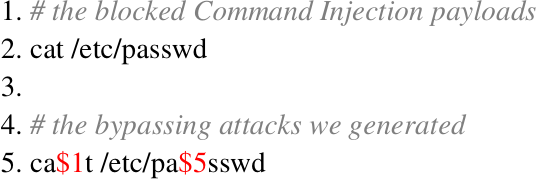}}
	\caption*{Code \thecodefigure: Case of generated payload of SQL injection (a), XSS (b), and command injection (c). Case a and b bypass the detection of Safedog, while Case c bypasses the detection of Modsecurity}
	\label{fig:cases}
\end{figure*}

\textbf{Generated XSS attack payload case.}
Code 1(b) shows a case of generated XSS payload that can bypass the detection Safedog.
After the user logs in, the browser keeps an authorization cookie.
By injecting a malicious payload into web pages, the attacker can gain elevated access privileges to cookies.
The case of the keyword is altered, and the space is replaced with `/' to evade Safedog.

\textbf{Generated command injection payload case.}
Code 1(c) shows a case of generated command injection payload that can bypass the detection of Modsecurity.
Most WAF creates rules to protect sensitive information, e.g., adding the \texttt{/etc/passwd} to the blacklist rule.
However, the bypassing payload is generated by inserting \$5 and \$1.
It can bypass the detection of WAF due to the fact that the combination of \$ and some characters are regarded as an empty substring by bash
and has no impact on the output.
This case demonstrates that \scheme is capable of generating a complicated malicious payload against the WAF.

\subsection{Comparison with SOTA Methods}
\label{sub:comparison}

We compared our system with the latest mutation-based method, WAF-A-MoLE~\cite{demetrio2020waf},
and state-of-the-art tool, SQLMap~\cite{sqlmap}.
WAF-A-MoLE employs an evolutionary algorithm to repeatedly alter test inputs through the random application of mutation operators, specifically targeting ML-based WAFs.
SQLMap is an open-source penetration testing tool targeting SQL, which is widely used to detect and exploit SQL injection vulnerabilities,
and we use its bypassing scripts to generate payloads.
Since SQL injection is a common attack and supported by WAF-A-MoLE and SQLMap,
we take the results of SQL injection for comparison.
We also used original malicious SQL injection payloads for evaluation as the baseline.
The two widely used WAFs, Modsecurity and Safedog are used for comparison.

\begin{table*}[!t]
	\renewcommand\arraystretch{0.85}
	\centering
	\scriptsize
	\caption{List of mutation operators in WAF-A-MoLE.}
	\begin{tabular}{@{}lll@{}}
		\hline
		\textbf{Operator}       & \textbf{Definition}                                                          & \textbf{Example}                                                                \\
		\hline
		Case swapping           & \texttt{CS(...a...B...)→...A...b... }                                        & \texttt{CS(admin' OR 1=1\#)→ ADmIn' oR 1=1 \# }                                 \\
		Whitespace substitution & \texttt{WS(...$k_1k_2$...)→...$k_1$␣$k_2$... }                               & \texttt{WS(admin' OR 1=1\#)→admin'\textbackslash{}n OR \textbackslash{}t 1=1\#} \\
		Comment injection       & \texttt{CI(..$k_1k_2$...)→...$k_1$/**/$k_2$...   }                           & \texttt{CI(admin' OR 1=1\#)→admin' /**/OR 1=1\#         }                       \\
		Comment rewriting       & \texttt{CR(.../*$s_0$*/...\#$s_1$)→.../*$s'_0$*/...\#$s'_1$ }                & \texttt{CR(admin' /**/OR 1=1\#)→admin' /*abc*/OR 1=1\#xyz    }                  \\
		Integer encoding        & \texttt{IE(...n...)→...0x${[}n{]}_{16}$}                                     & \texttt{IE(admin' OR 1=1\#)→admin' OR 0x1=1\#      }                            \\
		Operator swapping       & \texttt{OS(...$\oplus$...)→...$\boxplus$...(with$\oplus$$\equiv$$\boxplus$)} & \texttt{OS(admin' OR 1=1\#)→admin' OR 1 LIKE 1\#     }                          \\
		Logical invariant       & \texttt{(...e...)→...e AND $\top$...}                                        & \texttt{LI(admin'OR 1=1\#)→admin' OR 1=1 AND 2\textless{}\textgreater{}3\#  }   \\
		\hline
	\end{tabular}%
	\label{tab:mutation}
\end{table*}

\textbf{Mutation-based method (WAF-A-MoLE).}
We used the mutation operators in Table~\ref{tab:mutation} to alter the payload while keeping the semantics of the injected queries.
The mutation operators include case swapping, whitespace substitution, comment injection, comment rewriting,
integer encoding, operator swapping,
and logical invariant.
For example, the case swapping operator randomly changes the capitalization of the keywords in a payload.
Since SQL is case-insensitive, the semantics of the query is not affected.
The comment rewriting operator randomly modifies the content of a comment.
We applied these mutation operators to the original payloads to generate bypassing payloads.
We generated ten datasets with 1,000 payloads and evaluated the average of FAR.

\begin{figure}[!t]
	\centering
	\includegraphics[width=.7\linewidth]{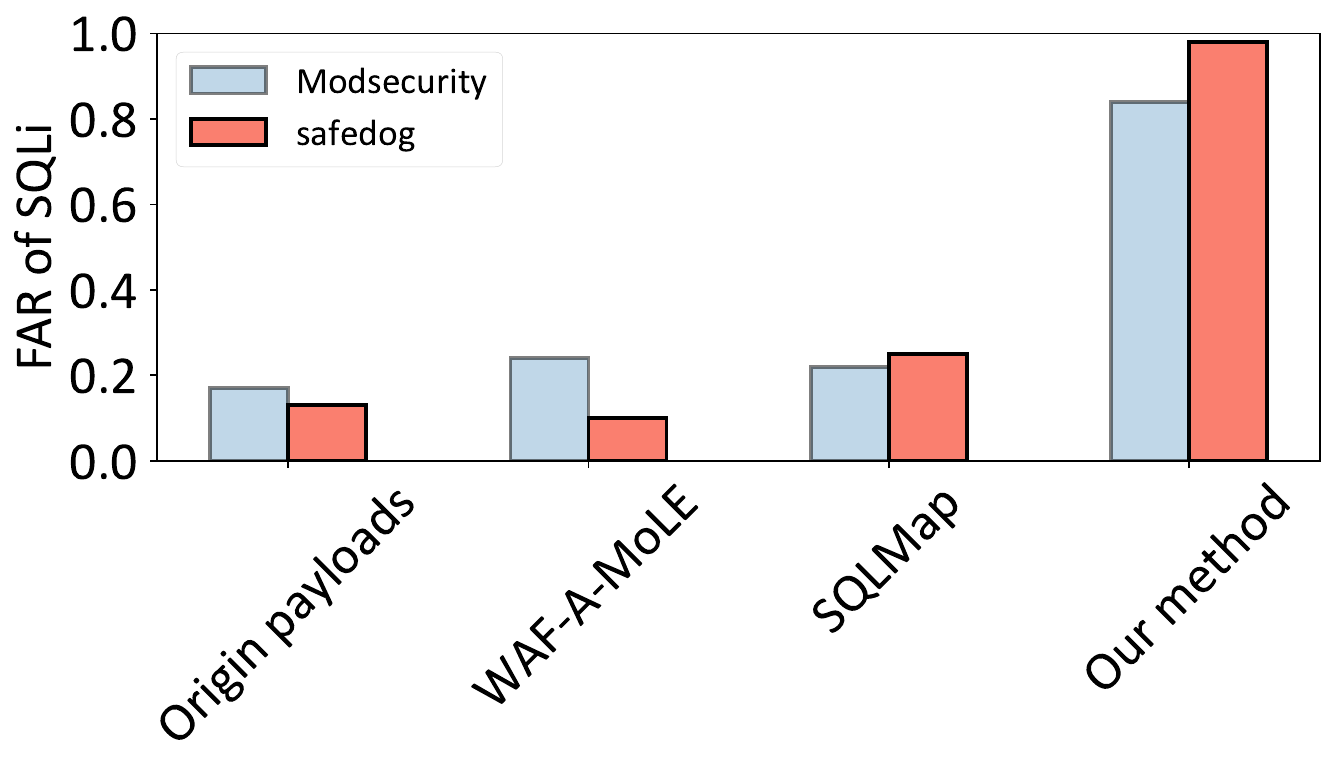}
	\caption{FAR of SQLi payloads against Modsecurity and safedog.}
	\label{fig:comp}
\end{figure}

\textbf{SQLMap.} SQLMap generates a diverse set of SQL injection payloads,
which covers all common SQL injection attack techniques.
The latest version (1.4.2.38) has 57 tamper scripts.
We removed some scripts related to encoding,
which are designed for specific backends and have a great impact on the accuracy of detecting malicious payloads.
We do the same operations as mutation-based schemes to evaluate mean FAR.

\textbf{Results.}
Fig.~\ref{fig:comp} shows FARs of SQL injection payloads against ModSecurity and safedog under original payloads,
WAF-A-MoLE, SQLMap, and our method.
Results show that these methods can generate SQL injection payloads to bypass the detection of ModSecurity and safedog,
suggesting that these two WAFs do not provide enough security.
For Modsecurity,
it can be seen that WAF-A-MoLE and SQLMap improve the FAR of generated payloads over the baseline,
but our method significantly improves the FAR up to 80\%,
which is more than 3 $\times$ of the other two methods.
As an example of ModSecurity,
WAF-A-MoLE and SQLMap achieve a FAR of 24\% and 22\%, respectively, while our method is with a FAR of 84\%.
For safedog, the results are similar, but the FAR of WAF-A-MoLE is smaller than the baseline.
We speculated that safedog may provide particular protection against mutation since the mutated payloads are different from benign SQL queries.
The mutated payloads can be easily detected by safedog, which indicates that the mutation scheme is not generalized for different WAFs.
Overall, the results demonstrate that \scheme generates malicious payloads with a higher FAR, significantly outperforming SQLMap and other methods.

Note that WAF-A-MoLE is designed for ML-based WAFs, applying mutation operators by prioritizing those that degrade ML model accuracy. However, the WAFs in our study, such as ModSecurity, are traditional rule-based WAFs. To ensure a fair evaluation, we adapted WAF-A-MoLE by using the same initial test inputs and applying mutation operators without the prioritization step. This generated diverse payloads, though it may not fully utilize WAF-A-MoLE's optimization techniques. Applying WAF-A-MoLE to non-ML-based WAFs may introduce a comparison bias. Our goal is to demonstrate the robustness and adaptability of our method across various WAF types, rather than to directly compare different tools.

\subsection{Generalization Ability of Shadow Model}

We evaluated the performance of generalizing the shadow model to other WAFs.
A generalized shadow model will reduce the shadow models' training costs for different WAFs.
The shadow model of Modsecurity and safedog were used for evaluation.
We fed the corrected payloads to the shadow models.
Modsecurity has three shadow models for SQL injection, XSS, and command injection, respectively.
Safedog has two shadow models.

Table~\ref{tab:universality} and~\ref{tab:universality_safedog} reports the TRR
when generalizing the shadow model of Modsecurity and Safedog to other WAFs, respectively.
The results show that the TRR of the generalized shadow model is similar to the original model.
For example,  when generalizing safedog shadow model to Modsecurity to detect SQL injection payloads, the original shadow model is 16\%, while the generalized shadow model is 14\%
We also speculated that the increase in TRR may be due to the shadow model is not specially designed for the target WAF that can block more attack payloads.
And the randomness of generating the payloads may also account for the increase and decrease of TRR.
Although the TRR increases or decreases, we think it is acceptable for enhancing the security of  WAF, and the shadow model can be generalized to other WAFs.

\begin{table*}[!t]
	\renewcommand\arraystretch{0.95}
	\caption{TRR of original shadow model. / TRR of generalizing the Modsecurity shadow model to other real WAFs.}
	\centering
	\scriptsize
	\scalebox{0.8}{
		\begin{tabular}{ccccccccccccccc}
			\hline
			\textbf{Attack type}                       &
			\textbf{naxsi}                             &
			\textbf{lua-resty-waf}                     &
			\textbf{safedog}                           &
			\textbf{xwaf}                              &
			\textbf{openwaf}                           &
			\textbf{verynginx}                         &
			\textbf{unixhot}                             \\
			\hline
			SQLi                                       &
			25\%  / 27\%                               &
			20\%         /{ 24\%}                      &
			2\%                               /	{ 2\%} &
			5\%                               /	{ 3\%} &
			3\%                               /	{ 7\%} &
			2\%                               /	{ 1\%} &
			{N/A}                                        \\
			XSS                                        &
			53\%                / { 65\%}              &
			59\%                 /{ 60\%}              &
			5\%                   / { 6\%}             &
			7\%                    / { 8\%}            &
			10\%                /{ 6\%}                &
			{N/A}                                      &
			3\%                  /{ 6\%}                 \\
			CI                                         &
			23\% /{ 31\%}                              &
			34\% / { 44\%}                             &
			{N/A}                                      &
			{N/A}                                      &
			{N/A}                                      &
			{N/A}                                      &
			{N/A}                                        \\ \hline
		\end{tabular}
	}
	\label{tab:universality}
\end{table*}

\begin{table*}[!t]
	\renewcommand\arraystretch{0.95}
	\scriptsize
	\caption{TRR of original shadow model. / TRR of generalizing safedog shadow model to other real WAFs.}
	\centering
	\scalebox{0.8}{
		\begin{tabular}{ccccccccccccccc}
			\hline
			{\textbf{\textbf{\textbf{Attack type}}}}   &
			{\textbf{\textbf{\textbf{Modsecurity}}}}   &
			{\textbf{\textbf{{naxsi}}}}                &
			{\textbf{\textbf{\textbf{lua-resty-waf}}}} &
			{\textbf{\textbf{\textbf{xwaf}}}}          &
			{\textbf{\textbf{\textbf{openwaf}}}}       &
			{\textbf{\textbf{\textbf{verynginx}}}}     &
			{\textbf{\textbf{\textbf{unixhot}}}}         \\
			\hline
			SQLi                                       &
			16\% / 			{ 14\%}                          &
			25\%   / { 27\%}                           &
			20\%     / { 42\%}                         &
			5\%        / { 4\%}                        &
			3\%          / { 4\%}                      &
			2\%            /  { 3\%}                   &
			{N/A}                                        \\
			XSS                                        &
			55\%                             / { 62\%} &
			53\%                              /{ 65\%} &
			59\%                              /{ 62\%} &
			7\%                               /{ 9\%}  &
			10\%                              /{ 5\%}  &
			{N/A}                                      &
			3\%              / { 7\%}                    \\ \hline
		\end{tabular}
	}
	\label{tab:universality_safedog}
\end{table*}

\subsection{Effectiveness of Signature Producing}
\label{sub:signature}
To evaluate effectiveness of our generated signatures, we measured TRR and TAR by applying these signatures to enhance WAF. We used bypassing adversarial payloads and benign payloads from original dataset. Integrating new rules into the existing WAF framework aims to enhance, not disrupt, the current rule set. To address concerns about misclassifying benign payloads, we focused on unaddressed threats and rigorously tested the new rules against synthetic and real-world traffic. We ensured the new rules/signatures effectively reject malicious payloads. Thus, payloads matching either previous rules or newly generated signatures were rejected.

Table~\ref{tab:signature} reports the TRR and FRR before and after applying the generated signature. Our method significantly improves TRR against malicious payloads, with all TRRs exceeding 90\% after using the signature. For instance, ModSecurity's TRR for SQL injection payloads increased from 16\% to 96\%. Additionally, the TAR of benign payloads is 100\%, indicating no benign payloads were rejected after applying the signatures. These results demonstrate the effectiveness of our signature producer in enhancing WAF security while maintaining high acceptance of benign requests. The 100\% TAR is due to the clear distinction between malicious and benign payloads, minimizing the chance of benign payloads being mistakenly identified as malicious.

\begin{table}[!t]
	\caption{Evaluation of the signature}
	\renewcommand\arraystretch{0.9}
	\centering
	\scriptsize
	\begin{tabular}{lcccr}
		\hline
		                                 &                                          & \multicolumn{2}{c}{{\textbf{TRR}}} &                                                     \\ \cline{3-4}
		\multirow{-2}{*}{{\textbf{WAF}}} & \multirow{-2}{*}{{\textbf{Attack type}}} & {\textbf{Before}}                  & {\textbf{After}} & \multirow{-2}{*}{{\textbf{FRR}}} \\
		\hline
		                                 & SQLi                                     & 16\%                               & 96\%             & 0                                \\
		                                 & XSS                                      & 55\%                               & 94\%             & 0                                \\

		\multirow{-3}{*}{Modsecurity}    & CI                                       & 33\%                               & 99\%             & 0                                \\
		\hline
		                                 & SQLi                                     & 25\%                               & 100\%            & 0                                \\

		                                 & XSS                                      & 53\%                               & 97\%             & 0                                \\
		\multirow{-3}{*}{naxsi}          & CI                                       & 23\%                               & 100\%            & 0                                \\
		\hline
		                                 & SQLi                                     & 20\%                               & 94\%             & 0                                \\
		                                 & XSS                                      & 59\%                               & 90\%             & 0                                \\
		\multirow{-3}{*}{lua-resty-waf}  & CI                                       & 34\%                               & 96\%             & 0                                \\
		\hline
		                                 & SQLi                                     & 2\%                                & 100\%            & 0                                \\
		\multirow{-2}{*}{safedog}        & XSS                                      & 5\%                                & 96\%             & 0                                \\
		\hline
		                                 & SQLi                                     & 5\%                                & 89\%             & 0                                \\
		\multirow{-2}{*}{x-waf}          & XSS                                      & 7\%                                & 98\%             & 0                                \\
		\hline
		                                 & SQLi                                     & 3\%                                & 92\%             & 0                                \\
		\multirow{-2}{*}{openwaf}        & XSS                                      & 10\%                               & 95\%             & 0                                \\
		\hline
		verynginx                        & SQLi                                     & 2\%                                & 97\%             & 0                                \\
		\hline
		unixhot                          & XSS                                      & 3\%                                & 97\%             & 0                                \\ \hline
	\end{tabular}
	\label{tab:signature}
\end{table}

\section{Related Work}
\label{sec:relwork}

\textbf{WAF security testing.}
As WAFs are increasingly used for security purposes to resist web attacks, ensuring the security of WAFs through effective testing methods becomes crucial. Previous methods, such as the decision tree-based method proposed by Dennis et al.~\cite{appelt2018machine}, require abstraction of attack syntax and have limitations in testing due to being derived only from summarized grammatical specifications. Additionally, these methods may only be useful for a single type of WAF. In contrast, our proposed method does not require summarizing grammatical specifications, and can be applied to various types of attacks and WAFs. While reinforcement learning-based vulnerability discovery methods~\cite{AmoueiRF22} have been proposed for WAF testing, they may only be effective for rule-based WAFs and cannot be used to test machine learning-based WAFs. Additionally, existing mutation generation schemes, such as that proposed by Demetrio et al.~\cite{demetrio2020waf}, are aimed only at WAFs based on machine learning and may not be effective for the more common black-box WAFs that use a combination of defense methods. Furthermore, existing schemes may require preset mutation operators and algorithms, making them complicated and only effective for specific WAFs. Our proposed method overcomes these limitations and has been tested on various types of WAFs with promising results.

\textbf{Adversarial example for text generation.}
Textual Adversarial Attacks: Beyond the visual domain, the last few years have witnessed growing interest in generating adversarial textual examples~\cite{gong2018adversarial,Li2019,AmoueiRF22,Gao2018a}.
The work of Gong et al.~\cite{gong2018adversarial} proposed a method to search for adversarial words in the embedding space, followed by reconstructing the adversarial sequence using nearest neighbor search and quantifying the quality of the adversarial text based on the distance of words. On the other hand, Gao et al.~\cite{Gao2018a} proposed to process the original input text directly. Li et al.~\cite{Li2019} proposed TextBugger, a general attack framework for generating adversarial text using five generation operations, including insertion, deletion, exchange, visual similarity replacement, and word vector space adjacent replacement. However, the generation operations in their work mainly address problems, e.g., sentence sentiment analysis, which is different from the generation of attacks. In contrast, our work uses a recurrent neural network to generate adversarial examples without defining generation operations.

We note that our work builds upon this foundational knowledge with a primary emphasis on WAFs, focusing on creating malicious payloads that can bypass the detection mechanisms.
Our approach carves a niche by offering an enhanced methodology specifically tailored for the nuances and challenges posed by WAFs.
It is not a departure from established research but rather an evolution, aiming to optimize detection and prevention mechanisms in the WAF context.

\textbf{Our method vs. fuzzing.}
Fuzzing involves sending a large volume of test inputs to a target program to uncover unexpected program behaviors and detect errors. The effectiveness and efficiency of the fuzzing process rely heavily on the quality of the test inputs, which can be generated using either mutation-based or generation-based approaches. In the generation-based approach, Wang et al.~\cite{Wang2017} proposed Skyfire, a data-driven seed generation method that leverages probabilistic context-sensitive grammar to generate seed inputs. In the mutation-based approach, You et al.~\cite{You2019} introduced ProFuzzer, a real-time detection technology that automatically recovers and understands the input fields that are critical for identifying vulnerabilities during the fuzzing process, and adapts mutation strategies intelligently to increase the chances of discovering zero-day vulnerabilities. Sablotny et al.~\cite{Sablotny2018} and Godefroid et al.~\cite{Godefroid2017} proposed using RNN to generate test inputs with a balance between fuzzing and structuring. In contrast to the conventional fuzzing approach, we employ adversarial examples to perform targeted and effective testing.

\section{Discussion}
\label{sec:dis}
Incorporating ML into WAFs provides a dynamic way to detect evolving threats, moving beyond the rigidity of traditional rule-based systems. However, this comes with challenges. The effectiveness of ML hinges on the quality of training data, demanding continuous updates as web threats evolve. Additionally, while deeper models excel in pattern recognition, they may introduce computational delays, conflicting with the real-time requirements of WAFs. This complexity also hampers model interpretability, contrasting with the transparent logic of rule-based systems. Periodic model retraining further demands resources, and the probabilistic nature of ML can lead to new false alarms. Such trade-offs between adaptability and operational demands explain the cautious approach of leading WAFs towards fully embracing ML solutions.

We have identified that the vulnerability of bypassing attacks originates from incomplete WAF security rules. As web application attacks continue to proliferate, more and more web applications are found to be vulnerable. While our system addresses many of these issues, there are some limitations and open problems that require further study. For instance, our system does not support DDoS attacks as the payload is typically harmless and cannot be blocked by WAF. Additionally, our system is highly dependent on the detected malicious datasets for model generation and benign datasets for shadow model training. If these datasets are poisoned, system performance will decline sharply.
Due to the unavailability of more realistic dataset, we evaluated the performance using previously collected dataset, as in other works.
In addition, the shadow models are individually limited on specific attacks, cannot not fully reflect the behavior of the target WAF.

We bolstered WAF effectiveness by crafting innovative payload generation techniques. When juxtaposed with established SOTA malicious datasets, our generated payloads showcased a heightened detection rate, signifying their meticulous design. Equally crucial was their real-world representativeness, ensuring WAFs are primed against genuine threats. By continuously evolving these payloads in response to the ever-changing threat landscape, our approach underscores the significance of adaptive security measures and stresses the importance of ongoing comparative evaluations with existing benchmarks.

To improve our system, we will focus on the following areas:
1) \emph{Improving the robustness of the shadow model.} A complex neural network can be used to train the shadow model, imitating the behavior of WAF with complicated rules.
2) \emph{Enhancing the effectiveness of generated payloads.} We will explore the use of other neural networks to enhance the effectiveness of the generated payloads.
3) \emph{Evaluating on more attacks.} It is necessary to extend our work to support more attacks, such as PHP injection (PHPi) and XML (XPath) injection (XMLi).
4) \emph{Broader Evaluation on realistic datasets.} It is also important to extend the evaluation of \scheme using more diverse and realistic datasets.

\section{Conclusion}
\label{sec:Conclusion}
We introduce \scheme, a black-box security testing framework leveraging learning techniques to identify and mitigate unknown malicious payloads in WAFs. By training a shadow model, \scheme generates both malicious and benign payloads, subsequently producing signatures for the former. Tested across eight real-world WAFs against three attack types, \scheme effectively uncovers numerous bypassing payloads and enhances WAF defense capabilities, demonstrating superior performance to current methods. As a potential future direction, we are looking forward to extending our method to improve the performance of various applications such as large language models~\cite{lin2023pushing,hu2024agentscomerge,hu2021lora,fang2024automated,lin2024splitlora}, distributed learning system~\cite{sun2024efficient,lin2024split,zhang2024fedac}, and secure systems~\cite{sun2024maglive,wu2024rethinking,xu2024sok,sun2024earpass,wu2024clad}.

\bibliographystyle{IEEEtran}
\bibliography{new}

\end{document}